\documentclass[prl,twocolumn,a4paper]{revtex4-1}
\usepackage{graphicx,amsmath,bm,natbib, dsfont}

\begin{document}

\title{Displacing entanglement back and forth between the micro and macro domains}
\date{\today}
\author{Natalia Bruno}
\affiliation{Group of Applied Physics, University of Geneva, CH-1211 Geneva 4, Switzerland}
\author{Anthony Martin}
\affiliation{Group of Applied Physics, University of Geneva, CH-1211 Geneva 4, Switzerland}
\author{Pavel Sekatski}
\affiliation{Group of Applied Physics, University of Geneva, CH-1211 Geneva 4, Switzerland}
\author{Nicolas Sangouard}
\affiliation{Group of Applied Physics, University of Geneva, CH-1211 Geneva 4, Switzerland}
\author{Rob Thew}
\affiliation{Group of Applied Physics, University of Geneva, CH-1211 Geneva 4, Switzerland}
\author{Nicolas Gisin}
\affiliation{Group of Applied Physics, University of Geneva, CH-1211 Geneva 4, Switzerland}

\begin{abstract}
Quantum theory is often presented as the theory describing the microscopic world, and admittedly, it has done this extremely well for decades. Nonetheless, the question of whether it applies at all scales and in particular at human scales remains open, despite considerable experimental effort. Here, we report on the displacement of quantum entanglement into the domain where it involves two macroscopically distinct states, i.e. two states characterised by a large enough number of photons to be seen, at least in principle, with our eyes and that could be distinguished using mere linear - coarse-grained - detectors with a high probability. Specifically, we start by the generation of entanglement between two spatially separated optical modes at the single photon level and subsequently displace one of these modes up to almost a thousand photons. To reliably check whether entanglement is preserved, the mode is re-displaced back to the single photon level and a well established entanglement measure, based on single photon detection, is performed. The ability to displace an entangled state from the micro to the macro domain and back again provides a fascinating tool to probe fundamental questions about quantum theory and holds potential for more applied problems such as quantum sensing.
\end{abstract}

\maketitle

Can quantum features, like entanglement, survive in the macro domain? This natural question has been with us since the inception of quantum theory. Nowadays, decoherence is widely accepted as one of the fundamental problems limiting the ability of macro systems to maintain quantum features~\cite{Zurek03}. As the size of a quantum system increases, it more and more intensively interacts with its surroundings, rapidly destroying its quantum properties. A few technologically demanding experiments, involving Rydberg atoms probing the electromagnetic field of a high-finess cavity~\cite{Brune96} or using a trapped ion interacting with engineered reservoirs~\cite{Monroe96, Turchette00}, have beautifully strengthened this idea. We also know that even if a system is sufficiently well de-coupled from the environment to hold its quantum features, they remain difficult to observe. One of the problems is that decoherence mechanisms inevitably increase the Hilbert space dimension, and even if a small amount of quantumness can survive, the system is no longer well described by the relatively simple theoretical models where measures are known for certifying quantum behaviour. This led for example, to lively debates~\cite{Sekatski09, Sekatski10, Spagnolo10, Spagnolo11, Demartini12} about the presence of entanglement in the experiment reported in Ref.~\cite{DeMartini08} where one photon from an entangled pair was phase-covariantly cloned. Measurement precision is another issue. Even if a macro system could be perfectly isolated from its environment, its quantum nature would require extremely precise measurements to be observed~\cite{Mermin80, Peres02}. These problems inspired the theoretical works presented in Refs.~\cite{Raeisi12, Sekatski12}, which proposed a means to prove the existence of entanglement in the macro domain by locally returning it back to the micro domain. As entanglement cannot increase through local operations, the detection of the micro entanglement proves the presence of entanglement in the macro domain. In particular, the approach of Ref.~\cite{Sekatski12} which is based on displacement in the phase space has several advantages. On the one hand, the existence of entanglement can be proven with well-established entanglement measures and witnesses operating on micro systems. On the other hand, the problem of the measurement precision is relegated to the displacement which can be performed very accurately.\\

In this letter, we report on the first experiment where entanglement is created in the micro domain and displaced back and forth between the micro and macro domains before being detected. The displacement is accurately controlled and allows for extending entanglement over unexplored regimes, ranging continuously from a single photon to almost a thousand. Under displacement, the entangled state becomes macroscopic not only because of the large mean photon number but also because its components could be distinguished with linear - coarse grained - detectors with a high probability. Paradoxically, demonstrating entanglement of the macro state requires precise measurements and our results provide evidence that the macro entanglement requires extreme phase resolution to be displaced back into the micro domain and subsequently detected.\\

\paragraph{Principle of the experiment}

\begin{figure*}[htbp]
{\includegraphics[width=1.9\columnwidth]{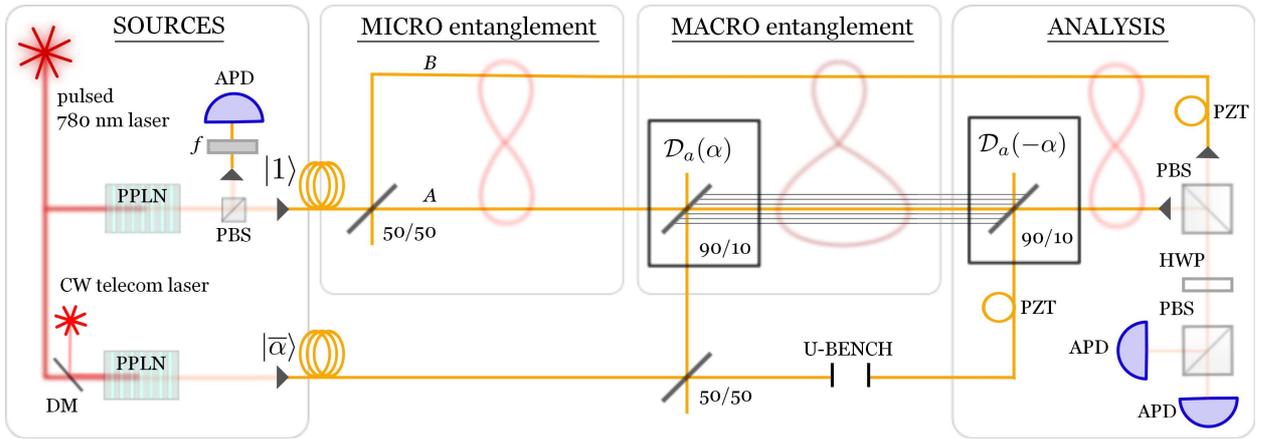}}
\caption{\textbf{Experimental setup}. SOURCES: A Ti:Sapphire laser pulsed in the picosecond regime with a 76\,MHz repetition rate at 780\,nm pumps two periodically poled lithium niobate (PPLN) bulk crystals.
In one crystal, pairs of photons are generated by type II spontaneous parametric down conversion and separated by a polarising beamsplitter (PBS). One photon is sent through a 25\,GHz filter (\textit{f}) at a central wavelength of 1563\,nm, and detected in a gated APD with $20\,\%$ detection efficiency. The second PPLN crystal is seeded by an additional CW laser at 1563\,nm, generating a local oscillator (by means of pulsed coherent state) via difference frequency generation, whose bandwidth is restricted by energy conservation to the bandwidth of the pump laser.
 MICRO entanglement: The heralded single photon (HSP) is sent to a fibre 50:50 beamsplitter, realising an entangled state between the two output modes $A$ and $B$. The figure-8 shapes represents the entanglement between the two modes. 
 MACRO entanglement: The mode $A$ is then combined with the local oscillator on a unbalanced beamsplitter (90:10), corresponding to a displacement operation on the HSP state. A set of black lines represents the fact that on this mode the state can be seen as macroscopic.
 ANALYSIS: The measurement setup consists in the application, on the mode $A$, of an inverse displacement operation $ \mathcal{D}_a(\alpha)^{-1} = \mathcal{D}_a(-\alpha)$ by means of another 90/10 fibre beamsplitter that closes a Mach-Zehnder interferometer. The adjustable U-bench allows one to accurately balance the path length difference for the experimental wavelength and to stabilise the phase on the side of the interference fringe from a reference laser (c.f. Annexe). In this interferometer, a piezo (PZT) is used to compensate the phase fluctuations. The PZT on the mode $B$ is used to observe single photon interference fringes. A pair of PBSs and a half waveplate (HWP) are used to recombine the modes $A$ and $B$ (HWP at $\pi/8$ ) or not (HWP at 0) to access $V$ or $p_{ij},$ respectively  (c.f. main text). The two outputs are detected by two APDs with $25\,\%$ detection efficiency, triggered by the detection of an idler photon in the HSP source.}
\label{fig1}
\end{figure*}

The experiment starts with the heralded creation of a path entangled state lying in the micro domain, see \figurename{~\ref{fig1}}. A nonlinear optical crystal is pumped by a pulsed laser in the picosecond regime to produce telecom photon pairs by means of spontaneous parametric down conversion. We set the pump intensity such that the probability of creating a single pair per run is very small. Thus, the detection of one idler photon heralds the creation of a single signal photon. By further placing a narrowband filter before the heralding detector, the signal photons are heralded in spectrally pure states with a bandwidth limited by the pump spectrum. To insure their spatial purity, they are coupled into a monomode fibre with an efficiency of $50\%$. The measurement of the second order auto-correlation $g^{(2)}(0)=1.9(1)$ unambiguously demonstrates the purity of the signal field~\cite{Christ11}. By sending the latter into a balanced beamsplitter, one obtains, leaving aside the loss, a maximally entangled state that describes the two output modes $A$ and $B$ sharing a single photon $\frac{1}{\sqrt{2}}\left(|1\rangle_A|0\rangle_B+|0\rangle_A|1\rangle_B\right).$ This path entangled state, known as single-photon entanglement, can be seen as the signature of the non-classical feature of the heralded signal photon~\cite{Asboth05, Solomon11}. \\

\begin{figure}
\begin{minipage}{\columnwidth}
\begin{tabular}{c}
a)\\
\includegraphics[width=1\columnwidth]{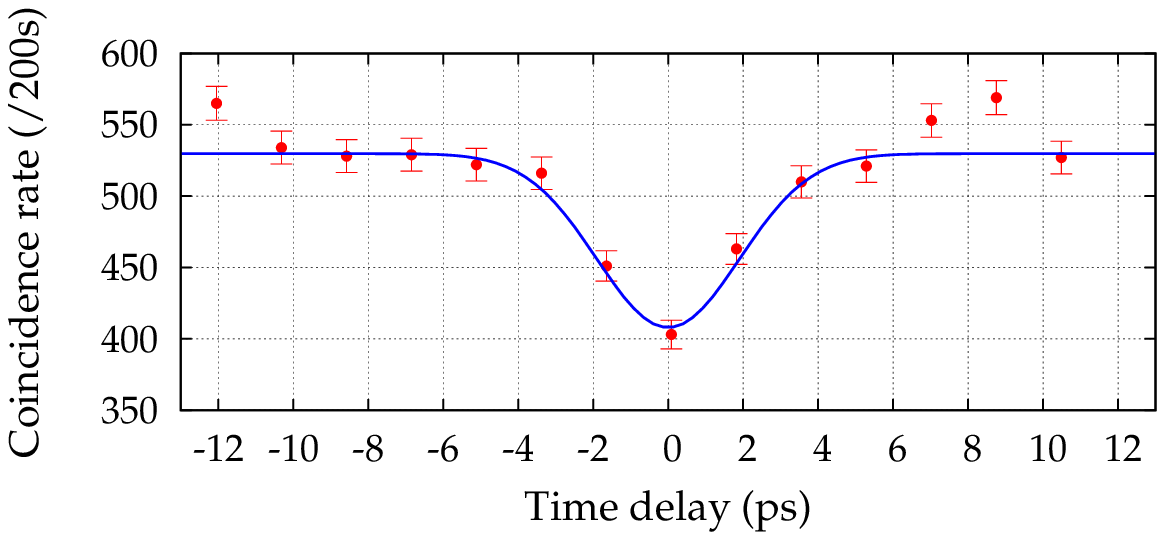}\\
b)\\
\includegraphics[width=1\columnwidth]{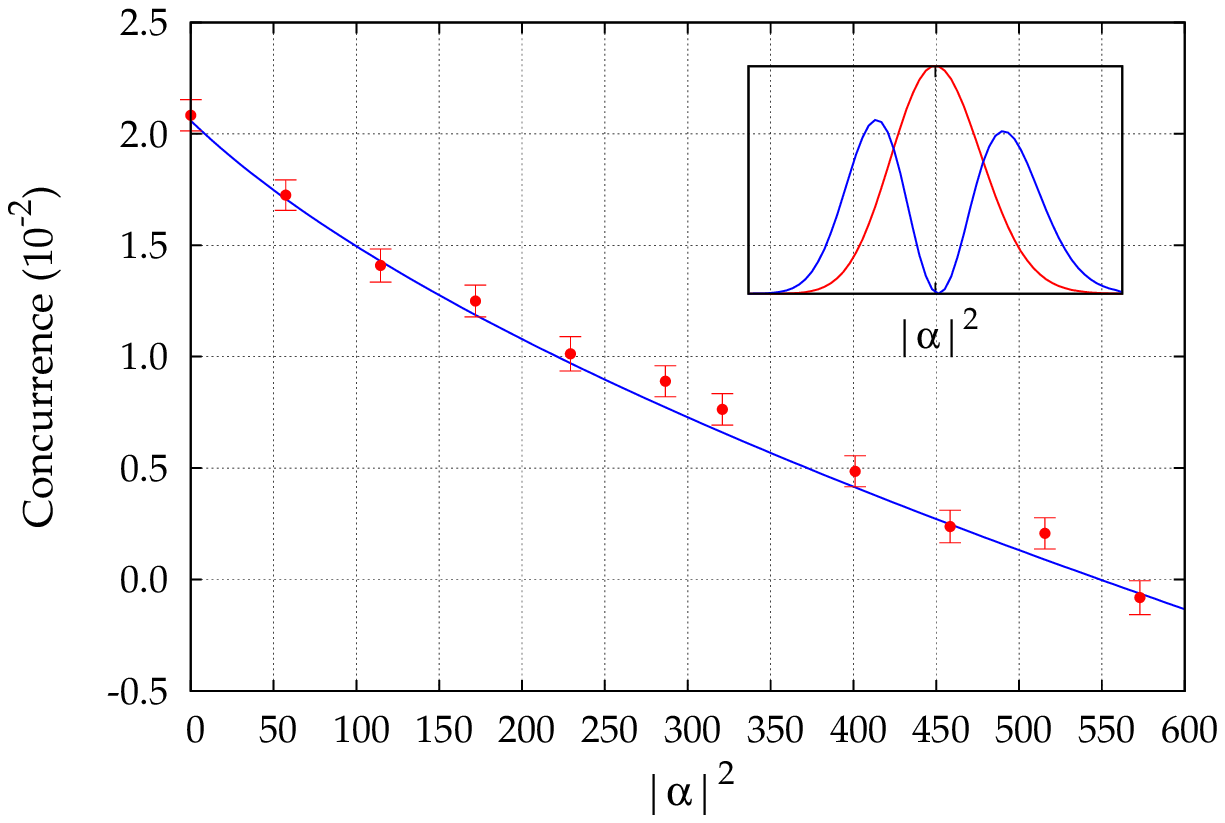}
\end{tabular}
\end{minipage}
\caption{{\textbf a.} HOM type interference between the heralded single photon in mode $A$ and the local oscillator that is used to displace it. This is obtained by varying the arrival time of the local oscillator onto the first 90:10 beam-splitter (c.f. \figurename{~\ref{fig1}}) and recording the twofold coincidences at the outputs of the beamsplitter. The experimental results (red dots) exhibits a dip with a visibility of $23 (4)\%$ {\textbf b.} The lower bound on the concurrence as a function of the mean photon number $|\alpha |^2$. The concurrence decreases when the photon number increases, as expected, but remains positive up to nearly 550 photons. The value of the underlying parameters ($V,$ $p_{00}$, $p_{01},$ $p_{10},$ and $p_{11}$) can be found in the Annexe. They are all based on raw counts, without subtracting dark counts and accidental coincidences. The full line is obtained from a theoretical model that uses independent measurements, c.f. main text. The inset shows the photon number probability distribution for $|\alpha\rangle$ (red) and $\mathcal{D}_a(\alpha)|1\rangle$ (blue).
}
\label{fig2}
\end{figure}

The second step consists in amplifying the mode $A$ by applying a unitary operation $\mathcal{D}_a(\alpha)$ corresponding to a displacement in the phase space. The latter is obtained by combining the mode $A$ and an intense local oscillator on a highly unbalanced beamsplitter~\cite{Paris96, Lvovsky02}. The physics behind the displacement is based on an interference process. Hence, the field $A$ and the local oscillator need to be indistinguishable. This is insured in practice by producing the local oscillator by means of a difference frequency generation (DFG), using an identical nonlinear crystal to the one used for the photon pair creation but stimulated by a cw telecom laser. The indistinguishability between the resulting local oscillator and the field $A$ is confirmed through a Hong-Ou-Mandel type interference~\cite{HOM87} whose dip, reported in \figurename{~\ref{fig2}}, has a visibility limited by the reflectivity of the beamsplitter and the photon statistics only. As such, after the displacement, the detection of an idler photon heralds the generation of an entangled state of the form
\begin{equation}
\label{micro_macro}
\frac{1}{\sqrt{2}}\left(\mathcal{D}_a(\alpha)|1\rangle_A|0\rangle_B+|\alpha\rangle_A|1\rangle_B\right).
\end{equation} 
$|\alpha\rangle_A$ results from the displacement of the vacuum. It follows a Poissonian photon number distribution with mean photon number $|\alpha|^2$ equal to the variance. The displacement also increases the mean photon number of the single photon state $|1\rangle,$ but it preserves its non-gaussian character. Specifically the state $\mathcal{D}_a(\alpha)|1\rangle_A$ is characterised by a photon number distribution with a mean photon number $|\alpha|^2+1$ and a variance $3 |\alpha|^2.$ The state (\ref{micro_macro}) thus describes entanglement between a microscopically populated mode $B$ and a mode $A$ whose mean population can be adjusted by tuning the intensity of the local oscillator. Remarkably, for large $|\alpha|^2,$ it involves a superposition of two components $|\alpha\rangle_A$ and $\mathcal{D}_a(\alpha)|1\rangle_A$ with largely different variances of the photon number distribution (see the inset in \figurename{~\ref{fig2}b}). In this regime, a mere binary detector that would only fire if the number of detected photons belongs to the interval $[|\alpha|^2 -|\alpha|, |\alpha|^2 + |\alpha|]$ would distinguish with a single shot the two components with a probability of 74\%. This makes the state (\ref{micro_macro}) analog to the famous Schroedinger cat state where the dead and alive components can be distinguished with very coarse-grained measurements. It also makes it very challenging to maintain since the sensitivity to phase instabilities is proportional to the variance of the photon number (for pure states). Note however, that this phase sensitivity may nonetheless provide a novel resource for quantum sensing~\cite{Sekatski12}.\\

In the last step of the experiment where we wish to probe whether entanglement survives as its size increases, the displaced state in mode $A$ undergoes another displacement $\mathcal{D}_a(\alpha)^{-1},$ which ideally returns it back to the single photon level. In practice, the de-amplification is realised in a similar manner as the amplification stage but inverting the phase of the local oscillator. The resulting fields can then be probed using single photon detectors to reveal heralded entanglement between the mode $A$ and $B.$ We emphasise that as the re-displacement is performed locally, it cannot increase the entanglement. A measure of the entanglement after the displacement thus provides a lower bound for the entanglement before the displacement, thus between the micro and macro states.\\

The state to be measured is described by a density matrix that includes noise and loss. To reveal entanglement, we use the tomographic approach based on single-photon detections presented in Ref.~\cite{Chou05} and successfully applied in many experiments~\cite{Chou05, Choi08, Usmani12, Lee11}. Specifically, from the measured values of the heralded probabilities $p_{mn}$ for detecting $m$ photons in mode $A$ and $n$ in mode $B$ $(m,n \in [0,1])$ and of the visibility $V$ of the interference obtained by combining the modes $A$ and $B$ on a balanced beamsplitter, a lower bound on the concurrence at the level of the detection is obtained through $C \geq V(p_{01}+p_{10})-2\sqrt{p_{00}p_{11}}.$ (n.b. the concurrence is a measure of entanglement ranging from 0 for separable states to 1 for maximally entangled qubit states). Let us mention straightaway that in order to maximise the observed concurrence, the interference visibility $V$ needs to be maximised and the probability for having one detection in each mode, $p_{11},$ has to be minimised. This can be done with ease at the single photon level, however, as the system size increases, this becomes increasingly difficult, as we shall see in the following.\\
 
The key ingredient that makes it possible to detect micro-macro entanglement in our scheme is to very accurately displace the macro component back to the single photon level. The typical size that can be achieved primarily depends on the precision with which the relative phase of the local oscillators is controlled, as small imperfections in the phase of the re-displacement inevitably add noise. The latter decreases the achievable interference visibility $V,$ and simultaneously increases $p_{11}$ leading rapidly to a zero concurrence. In our setup, the two local oscillators are produced by splitting the coherent state generated by the DFG process (see Supplementary Information). As presented in \figurename{~\ref{fig1}}, both displacements are made using 90:10 beamsplitters to reduce the single photon losses, and the second unbalanced beamsplitter closes a Mach-Zehnder interferometer. To accurately control and tune the relative phase between the two local oscillators, we actively stabilise the interferometer with a feedback loop locked on the side of an interference fringe using a tunable reference laser. We obtain an interference visibility of $99.985(2)\%(=1-\epsilon)$ which is essentially limited by the path length difference between the two arms ($\sim$20$\mu$m). This translates into an extinction ratio of $1.5(2)\times10^{-4}$, meaning that for $|\alpha|^2 = 6600(700),$ a noise of one photon is added into the detected mode. The theoretical model that we derived, which assumes that the relative phase of the two local oscillators is a stochastic variable (following a Gaussian distribution with variance $2\epsilon$) and that the source produces a two-mode squeezed state, and takes optical loss, detection efficiency, pair production probability into account, predicts a positive concurrence up to $|\alpha |^2=549,$ which is in excellent agreement to what we observed, as we shall see now.\\

\paragraph{Results} 
We first performed a series of measurements without displacement to verify that the mode $A$ and $B$ are indeed entangled when sharing a single photon.  The interference fringe obtained by combining the two modes on a 50:50 beamsplitter exhibits a visibility of $V=0.966(6).$ The conditional probabilities are measured to be $p_{00}=0.9719(1)$, $p_{01}=1.313(7)\times 10^{-2}$ $p_{10}=1.492(8)\times 10^{-2}$ and $p_{11}=1.0(2)\times 10^{-5}$ leading to $C \geq 0.0208(7)$. These measurements are then repeated for various displacements. \figurename{~\ref{fig2}} shows the resulting concurrence. The obtained values are in excellent agreement with our theoretical model that uses independent measurements of the transmission loss, detection efficiencies and pair creation probability. They show that the concurrence decreases as the state describing the mode $A$ becomes more macroscopic, and remains non-zero up to more than 500 photons. This strengthens the idea that one has to pay the price of increasing phase resolution to observe the quantum nature of a physical system as its size increases. The observed values are lower bounds on the amount of entanglement at the detection level and are mainly limited by optical loss. Factoring out the detector inefficiency and transmission loss yields an increase in the concurrence by more than one order of magnitude e.g. a lower bound of almost 20\% for the concurrence of the micro-macro state with $|\alpha|^2=100.$ Putting aside the loss before the displacement (the coupling inefficiency of the idler photon into the fibre), the concurrence would further increase to around 40\% for $|\alpha|^2=100.$\\

\paragraph{Conclusion}

We have reported an experimental observation of heralded entanglement involving large photon numbers and macroscopically distinguishable components. These results mark a significant increase in macroscopicity with respect to some previous efforts, in particular, the generation of optical Schroedinger kitten states, i.e. superpositions of coherent states with opposite phases involving a very few number of photons on average~\cite{Ourjoumtsev06, Neergaard06, Wakui07}. Furthermore, it also overcomes the problem faced by bright~\cite{Zhang00, Villar05, Keller08} and multi-mode ~\cite{Iskhakov12} squeezed states where the entangled components cannot be distinguished with coarse-grained measurements. So far, we have only considered entanglement at the level of one e-bit.  A fascinating perspective would be to store these photonic entangled states in atomic ensembles - the use of inherently multi-mode materials, like Rare-Earth doped crystal, would provide a particularly promising means to store several of these entangled states, thus further increasing the number of available e-bits. Similarly to NOON states~\cite{Afek10} or Squeezed states~\cite{LIGO11}, the phase sensitivity of the macro state under consideration may be useful for precision measurements, especially in the presence of loss, where alternative states have a limited usefulness. Returning to a more fundamental perspective, our results demonstrate that a physical system, here made with two clearly defined and well separated optical modes, can be displaced back and forth between the micro and macro domains while maintaining quantum entanglement. The principle could likely be generalised to demonstrate micro-macro entanglement obtained by any unitary operation, e.g. by cloning a photon belonging initially to an entangled pair, as proposed in Ref.~\cite{Raeisi12}. Our results also highlight the idea that although observing macro entanglement with coarse-grained measurements is very challenging, the creation of quantum macro systems can be straightforward. This suggests that quantumness is a concept that extends into our macro world and provides us with renewed motivation to look for quantum effects in Nature.\\

\paragraph{Acknowledgements}
We thank B. Sanguinetti and H. Zbinden for stimulating discussions and IDQ for the loan of one of the ID210 detectors. This work was supported in part by the EU project Q-Essence and the Swiss NCCR - Quantum Science and Technology (QSIT).

\bibliography{Bruno_micro_macro}

\clearpage

\section*{Annexe}

\textbf{A. Local oscillator source}
The coherent state source is composed of a 2\,cm PPLN crystal pumped by the same 780\,nm laser used in the heralded single photon (HSP) source (see main text) and seeded by a CW laser at 1563\,nm (idler wavelength in the HSP source).
Emission of pulsed light is stimulated by difference frequency generation (DFG). This process is governed by the energy conservation, which fixes the wavelength at 1557.5\,nm and limits the bandwidth to that of the pump laser. 
Having a common pump laser allows to prepare a HSP and a coherent state with same spectral properties.
A measurement of the second order autocorrelation function $g^2(0) = 1$ is used to confirm the Poissonian statistics of the state.

\vspace{1cm}
\noindent
\textbf{B. Indistinguishability study}

\begin{figure}
\includegraphics[width=0.8\columnwidth]{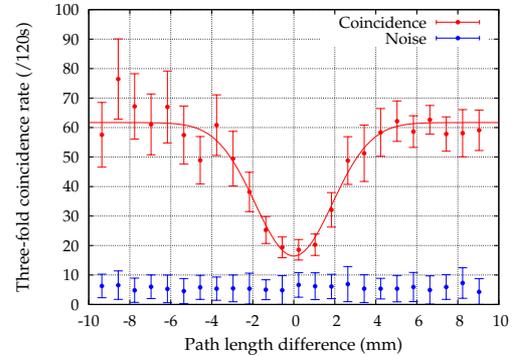}
\caption{Hong, Ou, and Mandel dip between HSP and LO on a balanced beam splitter. The visibility ($82(5)\%$) is in good agreement with what is expected taking in account the beam splitting ratio and the photon number statistics.}
\label{hom1}
\end{figure}

To prove the indistinguishability between the photons emitted by the two sources, HSP source and local oscillator, we realise a two photon interference experiment. The HOM dip maximal visibility is given by :
\begin{equation}
V_{max} = \dfrac{P_{1,1}}{P_{2,0}+P_{0,2}+\dfrac{r^2+t^2}{2rt}P_{1,1}},
\end{equation}
where $P_{i,j}$ represents the probability to have $i$ and $j$ photons at the input ports of the beamsplitter characterised by a transmission $t$ and a reflection $r$. To optimise this visibility, it is necessary to minimise the contribution $P_ {2,0}$ and $P_ {0,2}$ by adjusting the average number of photons in the two sources. In our experiment, the HSP source is characterised by a coupling efficiency close to 50\% and a mean number of photons of 0.01 per pulse. By setting the mean number of photon in the local oscillator at 0.05 photons per pulse, the maximal visibility is 80\% and 22\% for a 50:50 and a 90:10 couplers, respectively.  
As shown in \figurename{~\ref{hom1}}, and in \figurename{~\ref{fig2}}, we obtain experimentally visibilities of 82(5)\% and 23(4)\%, which proves the perfect indistinguishability between the photons from our two sources.  

\vspace{1cm}
\noindent
\textbf{C. Stabilisation}

In the last part of the setup (see "analysis" in \figurename{~\ref{fig1}) the state analysis stage is depicted. This part starts with a projection of the state into the single photon subspace, which is done in practice by applying an inverse displacement operator $ \mathcal{D}_a(-\alpha)$ on the mode $A$. 
In order to do this it is necessary to combine the mode $A$ with a local oscillator with the same properties as the first one and with opposite phase. For this purpose an interferometer with high visibility and phase stability is used. 
Due to the 4.5\,ps pulse duration of our local oscillator, the maximum visibility can be achieved with a perfectly balanced interferometer. Moreover, for the phase stability, it is necessary to lock the interferometer on a reference laser, but this cannot be done with 0 path length difference.  A good solution is the introduction of a different dispersion in one of the two arms. This is achieved by introducing a free space path ($\sim1$\,cm) in one arm, compensated (at the experimental wavelength) by  fibre path in the other arm. 
In this configuration, scanning the wavelength of the reference laser, we observe a full fringe every 400\,nm, from which we estimate a path length difference of 20\,$\mu$m.
Furthermore, we achieve visibilities of $99.98810(8)\%$ and $99.985(2)\%$ with a monochromatic laser and our local oscillator, respectively. The visibility in the pulse regime is mainly limited by the path length difference.
To set the phase of $\alpha$ in the re-displacement operation the interferometer is locked on the side of an interference fringe of the reference laser at 1460\,nm. 

\vspace{1cm}
\noindent
\textbf{D. Experimental results}

In \figurename{~\ref{vis_prob}} the raw data for the interference visibility and the probabilities $p_{m,n}$ (see main text) are shown. 

\begin{figure}
\begin{minipage}{\columnwidth}
\begin{tabular}{c}
a)\\
\includegraphics[width=.7\columnwidth]{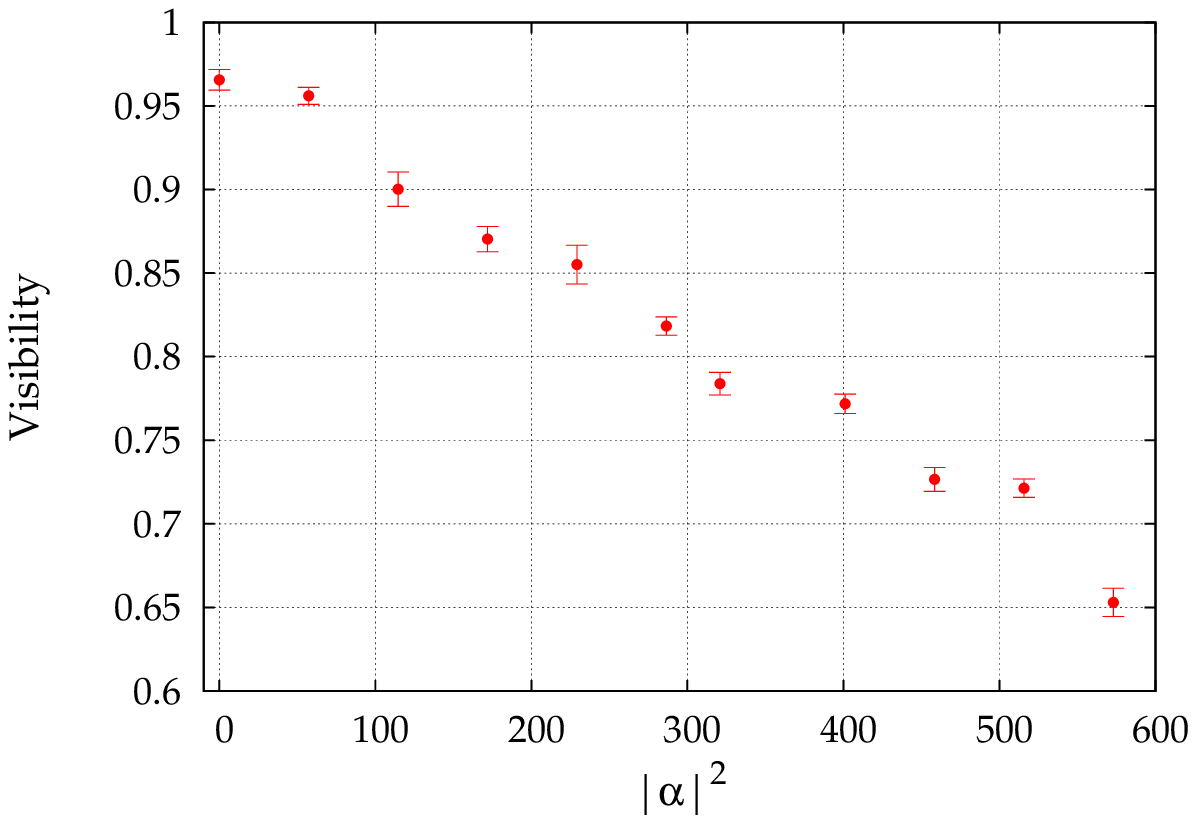}\\
b)\\
\includegraphics[width=.7\columnwidth]{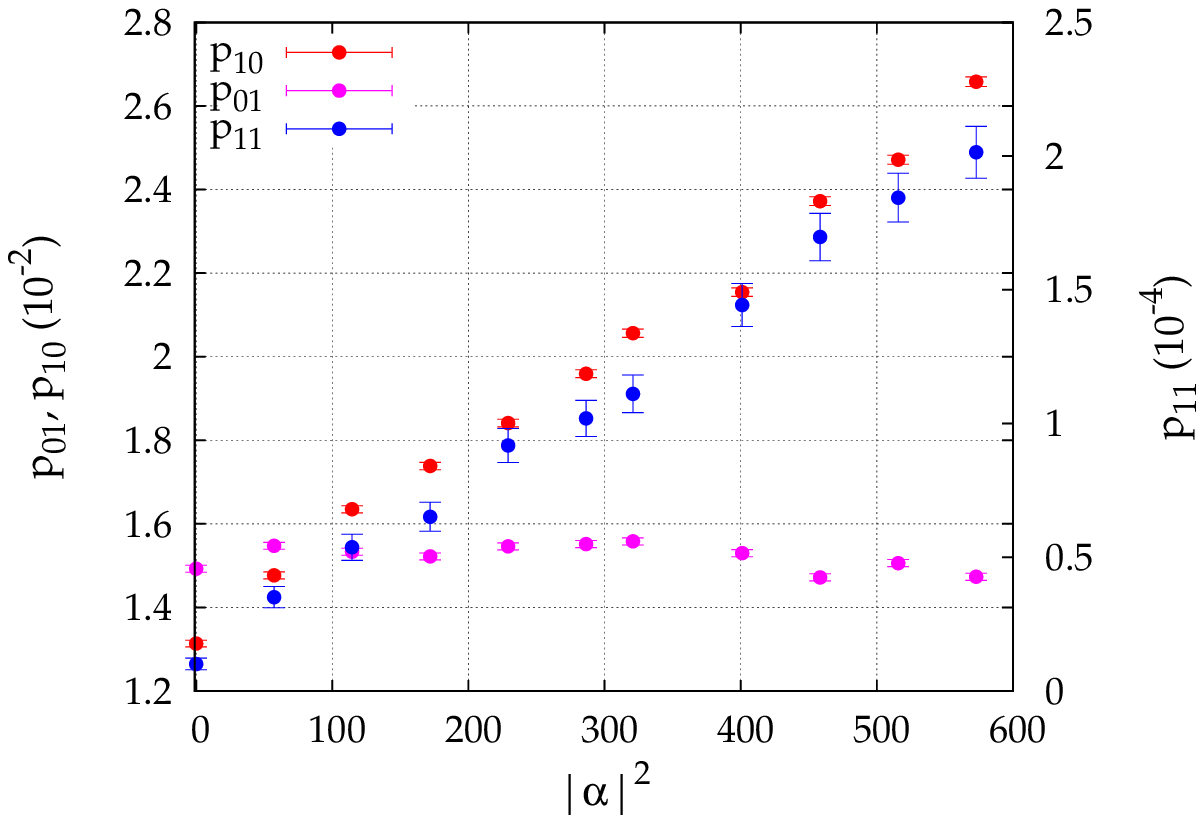}
\end{tabular}
\end{minipage}
\caption{\label{vis_prob}a) Single photon interference visibility as a function of the mean number of photons $\vert \alpha \vert^2 $ involved in the displacement. The drop in visibility comes from the additional noise introduced by the imperfect re-displacement operation.
b) Conditional probabilities $p_{m,n}$ as a function of the mean number of photons $\vert \alpha \vert^2 $ involved in the displacement. As expected, since the displacement is applied only on arm $A$ (see \figurename{~1}), $p_{01}$ is constant. On the other hand, $p_{10}$ and $p_{11}$ grow linearly with $\vert \alpha \vert^2 $, leading to a decreasing value for the concurrence (see \figurename{~\ref{fig2}}).}
\end{figure}

\end{document}